\documentclass[twoside,twocolumn,9pt]{article}
\usepackage{extsizes}
\usepackage[super,sort&compress,comma]{natbib} 
\usepackage[version=3]{mhchem}
\usepackage[left=1.5cm, right=1.5cm, top=1.785cm, bottom=2.0cm]{geometry}
\usepackage{balance}
\usepackage{mathptmx}
\usepackage{sectsty}
\usepackage{graphicx} 
\usepackage{lastpage}
\usepackage[format=plain,justification=justified,singlelinecheck=false,font={stretch=1.125,small,sf},labelfont=bf,labelsep=space]{caption}
\usepackage{float}
\usepackage{fancyhdr}
\usepackage{fnpos}
\usepackage[english]{babel}
\addto{\captionsenglish}{%
  
}
\usepackage{array}
\usepackage{droidsans}
\usepackage{charter}
\usepackage[T1]{fontenc}
\usepackage[usenames,dvipsnames]{xcolor}
\usepackage{setspace}
\usepackage[compact]{titlesec}
\usepackage[colorlinks=true,allcolors=blue]{hyperref}
\usepackage{bm}

\usepackage{epstopdf}

\usepackage{amssymb}

\definecolor{cream}{RGB}{222,217,201}

\begin{document}

\pagestyle{fancy}
\thispagestyle{plain}
\fancypagestyle{plain}{
\renewcommand{\headrulewidth}{0pt}
}

\makeFNbottom
\makeatletter
\renewcommand\LARGE{\@setfontsize\LARGE{15pt}{17}}
\renewcommand\Large{\@setfontsize\Large{12pt}{14}}
\renewcommand\large{\@setfontsize\large{10pt}{12}}
\renewcommand\footnotesize{\@setfontsize\footnotesize{7pt}{10}}
\makeatother

\renewcommand{\thefootnote}{\fnsymbol{footnote}}
\renewcommand\footnoterule{\vspace*{1pt}%
\color{cream}\hrule width 3.5in height 0.4pt \color{black}\vspace*{5pt}} 
\setcounter{secnumdepth}{5}

\makeatletter 
\renewcommand\@biblabel[1]{#1}            
\renewcommand\@makefntext[1]%
{\noindent\makebox[0pt][r]{\@thefnmark\,}#1}
\makeatother 
\renewcommand{\figurename}{\small{Fig.}~}
\sectionfont{\sffamily\Large}
\subsectionfont{\normalsize}
\subsubsectionfont{\bf}
\setstretch{1.125} 
\setlength{\skip\footins}{0.8cm}
\setlength{\footnotesep}{0.25cm}
\setlength{\jot}{10pt}
\titlespacing*{\section}{0pt}{4pt}{4pt}
\titlespacing*{\subsection}{0pt}{15pt}{1pt}

\fancyfoot{}
\fancyfoot[LO,RE]{\vspace{-7.1pt}\includegraphics[height=9pt]{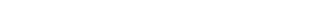}}
\fancyfoot[CO]{\vspace{-7.1pt}\hspace{13.2cm}\includegraphics{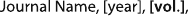}}
\fancyfoot[CE]{\vspace{-7.2pt}\hspace{-14.2cm}\includegraphics{head_foot/RF}}
\fancyfoot[RO]{\footnotesize{\sffamily{1--\pageref{LastPage} ~\textbar  \hspace{2pt}\thepage}}}
\fancyfoot[LE]{\footnotesize{\sffamily{\thepage~\textbar\hspace{3.45cm} 1--\pageref{LastPage}}}}
\fancyhead{}
\renewcommand{\headrulewidth}{0pt} 
\renewcommand{\footrulewidth}{0pt}
\setlength{\arrayrulewidth}{1pt}
\setlength{\columnsep}{6.5mm}
\setlength\bibsep{1pt}

\makeatletter 
\newlength{\figrulesep} 
\setlength{\figrulesep}{0.5\textfloatsep} 

\newcommand{\topfigrule}{\vspace*{-1pt}%
\noindent{\color{cream}\rule[-\figrulesep]{\columnwidth}{1.5pt}} }

\newcommand{\botfigrule}{\vspace*{-2pt}%
\noindent{\color{cream}\rule[\figrulesep]{\columnwidth}{1.5pt}} }

\newcommand{\dblfigrule}{\vspace*{-1pt}%
\noindent{\color{cream}\rule[-\figrulesep]{\textwidth}{1.5pt}} }

\makeatother

\renewcommand{\topfraction}{0.95}
\renewcommand{\bottomfraction}{0.95}
\renewcommand{\textfraction}{0.05}
\twocolumn[
  \begin{@twocolumnfalse}
{\includegraphics[height=30pt]{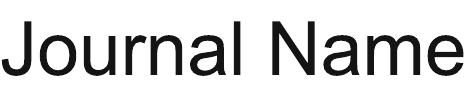}\hfill\raisebox{0pt}[0pt][0pt]{\includegraphics[height=55pt]{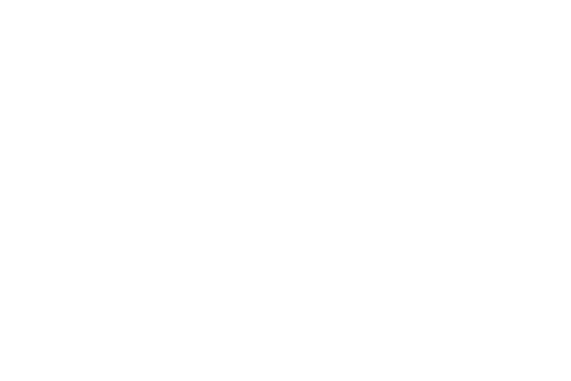}}\\[1ex]
\includegraphics[width=18.5cm]{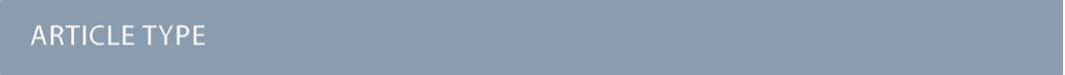}}\par
\vspace{1em}
\sffamily
\begin{tabular}{m{4.5cm} p{13.5cm} }

\includegraphics{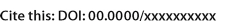} & 
\noindent\LARGE{\textbf{Models of 3D confluent tissue as under-constrained glasses}} \\
\vspace{0.3cm} & \vspace{0.3cm} \\

 & \noindent\large{Chengling Li,\textit{$^{a}$} Matthias Merkel,\textit{$^{b}$} and Daniel M. Sussman\textit{$^{a}$}$^\ast$} \\

\includegraphics{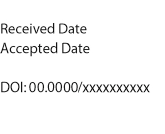} & \noindent\normalsize{%
  The dynamics of glassy materials slows down upon cooling, typically showing either Arrhenius or super-Arrhenius behavior.
	However, it was recently shown that 2D cell-based models for biological tissues can be continuously tuned between Arrhenius and \emph{sub}-Arrhenius dynamics.
	In previous work, using the 2D Voronoi model, we proposed that such atypical dynamical behavior could be a generic feature of the broad class of mechanically under-constrained materials.
	Our earlier study had left two important points open: (1) many 2D systems are affected by long-wavelength fluctuations and the 2D melting scenario, and (2) the 2D Voronoi model sits exactly at the isostatic point, making it a marginal case rather than a strictly under-constrained one.
	Both points complicate the interpretation of our 2D Voronoi model results and their generalization to other systems; to remedy this, here we use large-scale simulations to study the glassy behavior of the 3D extension of the Voronoi model.
	We first show that the structural relaxation time $\tau_\alpha$ of the 3D Voronoi model can be tuned between sub-Arrhenius and Arrhenius behavior, like the 2D Voronoi model.
    We then establish that the four-point susceptibility, the structure factor, and the model's mechanical properties all display trends consistent with the 2D Voronoi model.
	These results provide strong evidence that sub-Arrhenius glassy dynamics are a generic feature of under-constrained materials across dimensions.
    Our work thus broadens the class of disordered materials known to have highly unusual glassy phenomenology.
	} \\

\end{tabular}

 \end{@twocolumnfalse} \vspace{0.6cm}

  ]

\renewcommand*\rmdefault{bch}\normalfont\upshape
\rmfamily
\vspace{-1cm}


\footnotetext{\textit{$^{a}$~Department of Physics, Emory University, Atlanta, Georgia 30322, USA.}}
\footnotetext{\textit{$^{b}$~Aix Marseille Univ, Universit\'e de Toulon, CNRS, CPT (UMR 7332), Turing Centre for Living systems, Marseille, France}}
\footnotetext{\textit{$^\ast$~Email: daniel.m.sussman@emory.edu}}




\section{Introduction}

Understanding the origin of glassy dynamics remains one of the central challenges in condensed matter physics.
Upon cooling or compression, a wide variety of liquids and soft materials exhibit a dramatic slowdown of relaxation, often accompanied by increasing dynamical heterogeneity and the emergence of an intermediate-time plateau in time correlation functions \cite{cavagna2009supercooled,debenedetti2001supercooled,ediger1996supercooled,kob1995testing}.
Essentially all glasses display super-Arrhenius and Arrhenius temperature dependence of the relaxation time, and are classified as fragile and strong glassformers, respectively \cite{ediger1996supercooled}.
However, the universality of this trend has recently been challenged by the discovery of models that exhibit qualitatively different, sub-Arrhenius dynamics \cite{ciarella2019understanding,sussman2018anomalous}.
Exploring such departures is essential to build a more complete picture of glass formation and of the possible diversity of disordered dynamics.

Often, the glassy behaviors of materials can be closely linked to their mechanical properties.
In the zero-temperature limit, the mechanical properties of a system can be classified using the Maxwell constraint-counting criterion, which distinguishes three regimes.
Systems that are \emph{over-constrained} possess more constraints than degrees of freedom and are typically rigid.
In contrast, \emph{under-constrained} systems have fewer constraints than degrees of freedom and are mechanically unstable in the absence of additional interactions.
Between these two limits lie \emph{isostatic} systems, where the number of constraints and degrees of freedom are balanced, leading to marginal stability and weak solid-like behavior.
The role of a glass' mechanical properties in determining its dynamics has been the subject of much study \cite{dyre2012instantaneous}, and the extent to which a system's zero-temperature athermal mechanics dictates its finite-temperature dynamics remains a subject of active debate.

One system where atypical glassy dynamics has been linked to mechanical properties is the cell-based 2D Voronoi model for biological tissues.
This model has been shown to exhibit atypical sub-Arrhenius glassy dynamics \cite{sussman2018anomalous}, the extent of which can be tuned continuously \cite{ansell2025tunable}.
In recent work, we linked this dynamical behavior to this model's unusual, ``entropic'', elasticity, i.e.\ a strong increase of the elastic moduli with temperature \cite{li2025connecting}.
We further argued that such mechanical behavior is expected for under-constrained systems \cite{lee2024generic,Lee2024a}, which include many of the cell-based tissue models \cite{Merkel2019}.
We thus proposed that the tunable sub-Arrhenius behavior may be a generic property of under-constrained systems \cite{li2025connecting}.

While compelling, the interpretation of these findings is complicated by two aspects.
First, the model is two dimensional, and it is well-understood that long wavelength fluctuations play a vital role in two dimensions \cite{flenner2015fundamental}, complicating the analysis of quantities like the mean-squared displacement and self-intermediate scattering function~\cite{vivek2017long, illing2017mermin}.
Moreover, the two-dimensional melting scenario introduces additional subtleties through the possible emergence of a hexatic phase \cite{halperin1978theory, li2018role}, restricting the parameter regions where disordered dynamics can be studied.
Second, the 2D Voronoi model sits precisely at the isostatic constraint-counting transition, with equally balanced degrees of freedom and energetic constraints.
Consequently, while its athermal mechanical properties still reflect its proximity to under-constrained tissue models, it does not show an athermal floppy-rigid transition like those other models \cite{sussman2018no}.

To test whether the anomalous dynamical properties observed in the 2D Voronoi model are found in truly under-constrained materials, free from the confounding effects of hexatic ordering transitions, we study the 3D Voronoi model, which is extensively under-constrained \cite{merkel2018geometrically}.
As a consequence, like many cell-based tissue models \cite{bi2015density,sussman2018no,Merkel2019,Kim2024}, the athermal 3D Voronoi model exhibits a rigidity transition that is tuned by a target cell surface area $s_0$: at zero temperature, the model is rigid for $s_0\lesssim 5.4$ and floppy for $s_0\gtrsim 5.4$ \cite{merkel2018geometrically}.

Below, we systematically characterize the 3D Voronoi model at finite temperature, including its dynamical behavior, structural properties, and mechanical response across a range of $s_0$ values.
We show that the relaxation time exhibits sub-Arrhenius behavior whose extent increases with $s_0$, directly analogous to what has been observed in two dimensions.
The self-intermediate scattering function displays universal scaling when rescaled by the $\alpha$-relaxation time, and the growth of the four-point susceptibility is strongly suppressed at large $s_0$.
The static structure factor reveals a weak but systematic temperature dependence of its first peak, reminiscent of vitrimeric systems in which static correlations are tied to anomalous relaxation.
Underpinning these dynamical anomalies, we find that the isotropic tension and shear modulus exhibit a monotonic increase with temperature.
The isotropic tension also has a crossover from $T^{1/2}$ to $T^1$ scaling, consistent with theoretical predictions for generic under-constrained systems \cite{lee2024generic,Lee2024a}.
Together, these results establish that the unconventional glassy dynamics and mechanics of the Voronoi model persist in three dimensions.
This further supports the idea that sub-Arrhenius behavior is a generic property of under-constrained systems.

\section{Methods}

The 3D Voronoi model describes a biological tissue densely packed with $N$ cells \cite{merkel2018geometrically}. A given state of the model is defined by a set of cell center positions $\bm{r}_i$ in 3D space. Cell shapes are defined by a Voronoi tiling of these centers. 
Forces on cell positions, $\bm{f}_i=-\partial e/\partial\bm{r}_i$, are defined using an effective energy $e$, which reads in dimensionless form:
\begin{equation}\label{eq:energyFunctional}
	e=\sum_{i=1}^N\left[(v_i-1)^2 + k_{s,i}(s_i-s_0)^2\right].
\end{equation}
The sum is over all cells $i$, the variables $s_i$ and $v_i$ denote cell surface areas and volumes, respectively, $s_0$ is the target surface area, and $k_{s,i}$ is a surface area modulus.
We choose the unit of length to be the cube root of the average cell volume, which we furthermore set to unity.
This energy is analogous to how the 2D Voronoi model is defined \cite{bi2016motility}, with harmonic constraints on the cell perimeter and cross-sectional area promoted to constraints on the cell surface area and volume.
Like in many cell-based tissue models, the form of the energy is motivated by a low-order expansion of geometric cell properties in terms of target values \cite{Farhadifar2007,Staple2010,kim2018universal}. 
Note that this energy does not correspond to central force interactions, but rather many-body forces that couple the geometry and topology of the cellular material, which together generate the rigidity of the simulated tissue \cite{bi2015density, merkel2018geometrically, sussman2018no, sussman2018soft}.
We simulate this model using 3D periodic boundary conditions in a system with total volume $N$.

Initial work studying the 3D Voronoi model focused on the athermal limit of monodisperse systems\cite{merkel2018geometrically}, but just as in particulate settings there are model parameter regimes in which monodisperse Voronoi packings have a tendency to crystallize.
To maintain our focus on the disordered regime, we use a bidisperse mixture of cells whose surface area stiffness $k_s$ varies with cell type.
Here we choose a $80:20$ mixture of cell types $A$ and $B$ with $k_{s,A}/k_{s,B} = 1:0.1$.
All cells have the same $s_0$, and have equal masses (which we take to be unity).
We note that we also explored bidisperse mixtures with identical $k_s$ and cell-type-dependent $s_0$ or cell-type-dependent target cell volume, and we found qualitatively identical results.
However, we found that stiffness bidispersity let us explore a somewhat broader range of parameter space without observing crystallization.

To study finite-temperature equilibrium properties of this model, we perform standard NVT simulations \cite{frenkel2023understanding,martyna1996explicit}, with temperature implemented via a Nos{\'e}-Hoover thermostat to ensure proper sampling of the canonical ensemble.
After verifying that our conclusions are not qualitatively affected by changing the scale of the system, we have focused the bulk of our numerical work on simulations of $N=512$ cells, using an integration time step of $dt=10^{-2}\tau$ (where $\tau$ is the standard dimensionless simulation unit of time for this model).
We varied $s_0$ and temperature $T$, and obtained 10 equilibrium trajectories for each $(s_0,T)$ point.
To this end, for any given value of $s_0$, we first equilibrated at $T_\mathrm{init}=0.1$ (for which the characteristic relaxation time $\tau_\alpha$ is on the order of $1-10\tau$).
We then saved configurations after every $10^4\tau$, which is large enough to ensure that the configurations are independent from each other. These then serve as initial configurations for the simulations with the actual temperature $T$.
From these initial configurations, we apply an initial thermalization period at the target temperature $T$. We use preliminary simulations to measure $\tau_\alpha$ at the target temperature, thermalize for at least $10\tau_\alpha$, and then begin recording trajectory data for another $10\tau_\alpha$, which we analyze to collect the statistics reported below.

Given the high computational cost of simulating this model, our results primarily focus on four representative $s_0$ values: $s_0\in\{5.5,5.425,5.39,5.37\}$; these values correspond to surface areas well above that of a densely packed crystalline configuration \cite{merkel2018geometrically}, and are represented by filled symbols in all plots.
For each $s_0$, we select temperatures such that the resulting structural relaxation times span from $10\tau$ to $10^4\tau$.
In order to further study the $s_0$-dependence of the mechanical properties, we complement these simulations with additional ones with $s_0\in\{5.475,5.45,5.4,5.35,5.32,5.31,5.3,5.29,5.26,5.2\}$ but over a more limited temperature range corresponding to $\tau_\alpha \sim 10 \dots 10^3\tau$; these data -- which include surface areas below that of the densest known crystalline packings  -- are represented by hollow symbols in all plots.

\section{Results}

\begin{figure}
	\centering
	\includegraphics[width=\linewidth]{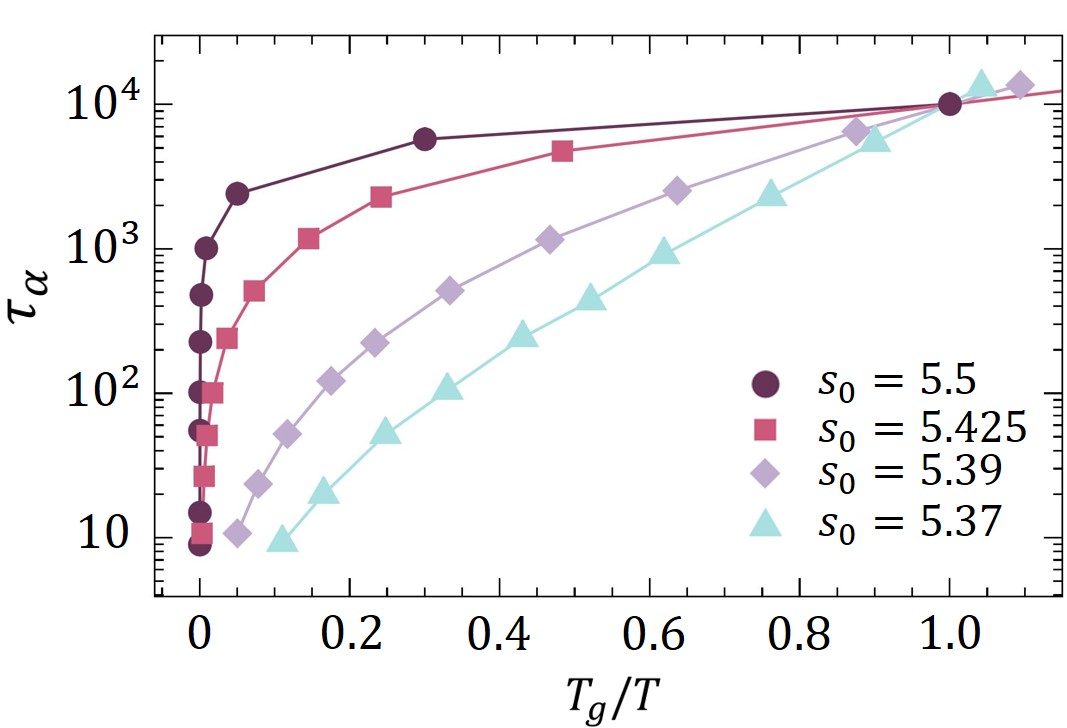}
	\caption{
		\textbf{The 3D Voronoi model exhibits tunable sub-Arrhenius relaxation dynamics.}
		Angell plot of the $\alpha$-relaxation time, $\tau_\alpha$, for four values of $s_0$.
		Each curve represents a different $s_0$ from high to low (dark red to light blue), and for each curve, $T_g$ is determined separately as the temperature where $\tau_\alpha=10^4$.
	}
	\label{fig:angelplothmc}
\end{figure}

\subsection{Dynamical properties}
We observe tunable sub-Arrhenius behavior also in the 3D Voronoi model.
\autoref{fig:angelplothmc} shows the dependency of the relaxation time $\tau_\alpha$ on temperature $T$ in an Angell plot for several values of the control parameter $s_0$; these relaxation times were extracted from the behavior of the self-intermediate scattering function, $F_s(t)$, as detailed in Appendix A.
In all cases, this representation reveals sub-Arrhenius (i.e.\ concave) behavior.
This behavior depends systematically on $s_0$: systems with larger $s_0$ exhibit more pronounced sub-Arrhenius scaling, while decreasing $s_0$ drives the dynamics closer to an Arrhenius form.
This result demonstrates that the 3D Voronoi model inherits the same key feature identified previously in two dimensions: unusual glassy dynamics whose details are tuned continuously via the target shape parameter of the model \cite{ansell2025tunable}.

\begin{figure}[t]
	\centering
	\includegraphics[width=\linewidth]{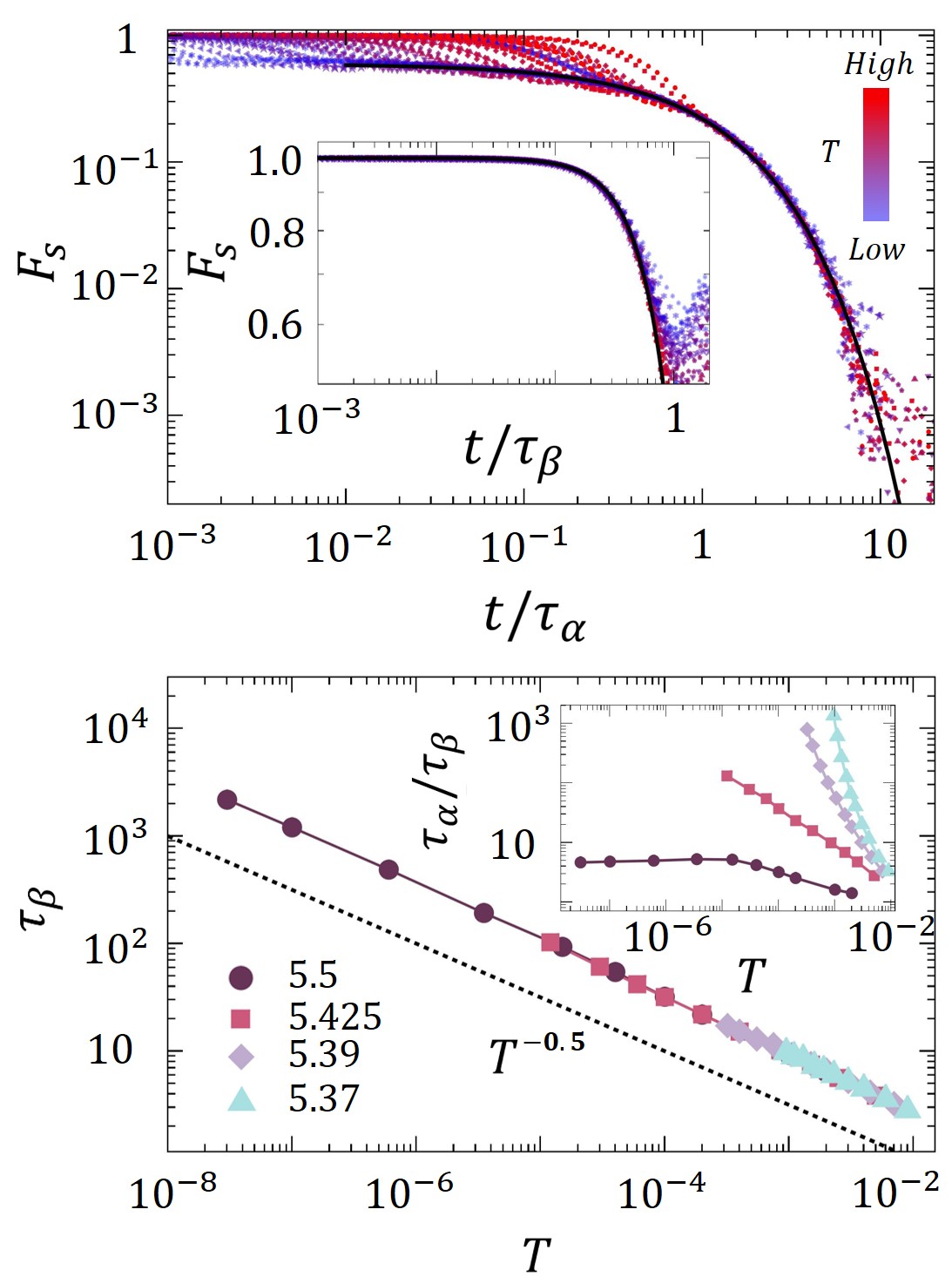}
	\caption{
		\textbf{$\alpha$- and $\beta$-relaxation from the self-intermediate scattering function.}
        (\textbf{Top})  $F_s(t)$ from the trajectories across all $(s_0,T)$ parameter pairs, plotted with time scaled by $\tau_\alpha$, showing collapse of all curves in the $\alpha$-relaxation regime. The black line is the fitting result. The inset shows $F_s(t)$ with the time axis scaled by $\tau_\beta$.
        (\textbf{Bottom}) The measured $\tau_\beta$ as a function of $T$ for several values of $s_0$; the dotted black line is a guide to the eye with slope $-1/2$. The inset shows the ratio of the $\alpha$ and $\beta$ relaxation times as a function of temperature; a clear cross-over is observed from a regime at large $s_0$ in which the time scales have the same temperature dependence to a more normal regime at low $s_0$ in which $\tau_\alpha$ grows much more rapidly as the temperature is decreased.
        }
	\label{fig:fs}
\end{figure}

The top panel of \autoref{fig:fs} shows the self-intermediate scattering function, $F_s(k,t)$, evaluated at the wavevector corresponding to the first peak ($k=k_{max}$), of the static structure factor for different values of $s_0$ and $T$. 
We choose a representation in which the time axis is scaled by our measured $\tau_\alpha$ for each $\{s_0,T\}$ pair. 
Remarkably, all curves from different $(s_0,T)$ pairs collapse onto a single master curve in the $\alpha$-relaxation regime.
This collapse demonstrates that, despite quantitative differences in relaxation time, the underlying $\alpha$-relaxation process is universal across the range of $s_0$ values and temperatures studied.

This representation also makes clear, though, that there are systematic shifts in how the plateau in the two-step relaxation process is approached. 
We separately measure this initial relaxation time, $\tau_\beta$,  by fitting the initial relaxation of $F_s$ to a stretched exponential decay. We show in the inset that this process, too, shares a common functional form across state points.
We thus plot the behavior of this faster relaxation process for a range of $\{s_0,T\}$.
As shown in the bottom panel of \autoref{fig:fs}, we find that in this model the beta relaxation is \emph{much} more sensitive to temperature than either standard particulate glass formers or the 2D Voronoi model, with a clear $1/\sqrt{T}$ dependence across all $s_0$ values studied.
We are not aware of other glassforming models that show such a range of behavior in the competition between the alpha and beta relaxation process scales.

\begin{figure}[t]
	\centering
	\includegraphics[width=\linewidth]{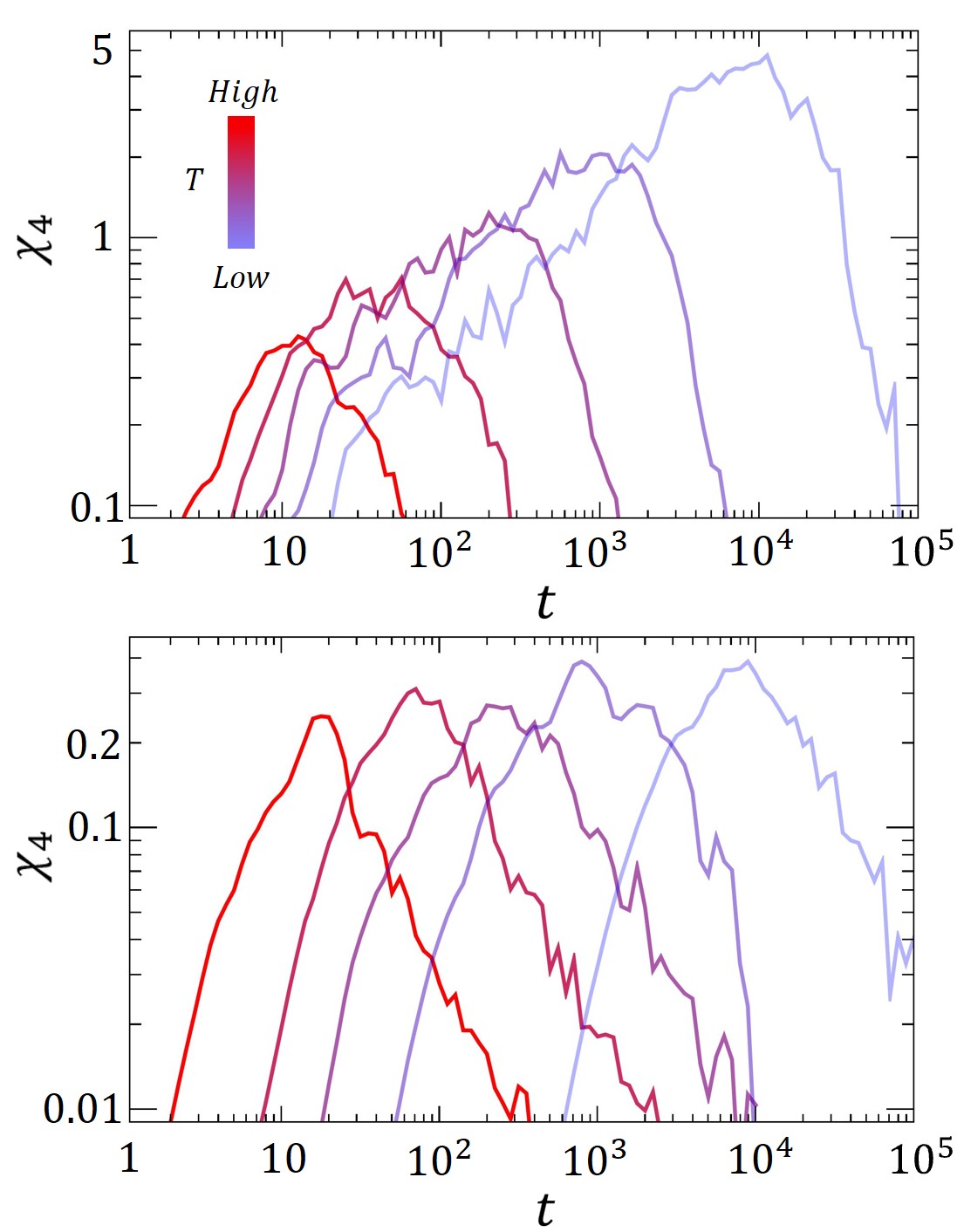}
	\caption{
		\textbf{The temperature dependence of dynamical heterogeneity is controlled by $s_0$.}
		Four-point dynamical susceptibility $\chi_4(t)$ for $s_0=5.39$ (Top) and $s_0=5.50$ (Bottom) at different temperatures (temperature decreases from dark red to light blue).
	}
	\label{fig:chi4}
\end{figure}

\autoref{fig:chi4} shows the four-point dynamical susceptibility $\chi_4(t)$ at different temperatures for two representative values of the control parameter, $s_0=5.5$ (bottom) and $s_0=5.39$ (top).
The peak height of $\chi_4(t)$ is often interpreted as a measure for the magnitude of dynamical heterogeneities in the system \cite{berthier2011overview}.
For $s_0=5.5$, the peak height of $\chi_4(t)$ is essentially independent of temperature, but for $s_0=5.39$ the peak height of $\chi_4(t)$ increases systematically upon cooling.
This behavior closely parallels previous observations in two dimensions: at higher $s_0$, dynamical heterogeneity is strongly suppressed, while decreasing $s_0$ restores the conventional growth of heterogeneous dynamics upon cooling \cite{ansell2025tunable}.
The ability of the Voronoi model to tune not only the global relaxation timescale but also the degree of dynamical heterogeneity highlights the unusual character of glass formation in this system.

Taken together, the behaviors of structural relaxation and dynamical heterogeneities in the 3D Voronoi model are very close to what we had observed in 2D \cite{li2025connecting, ansell2025tunable}.
This suggests that they are robust properties of the Voronoi model, rather than an artifact of dimensionality.

\begin{figure}[t]
	\centering
	\includegraphics[width=\linewidth]{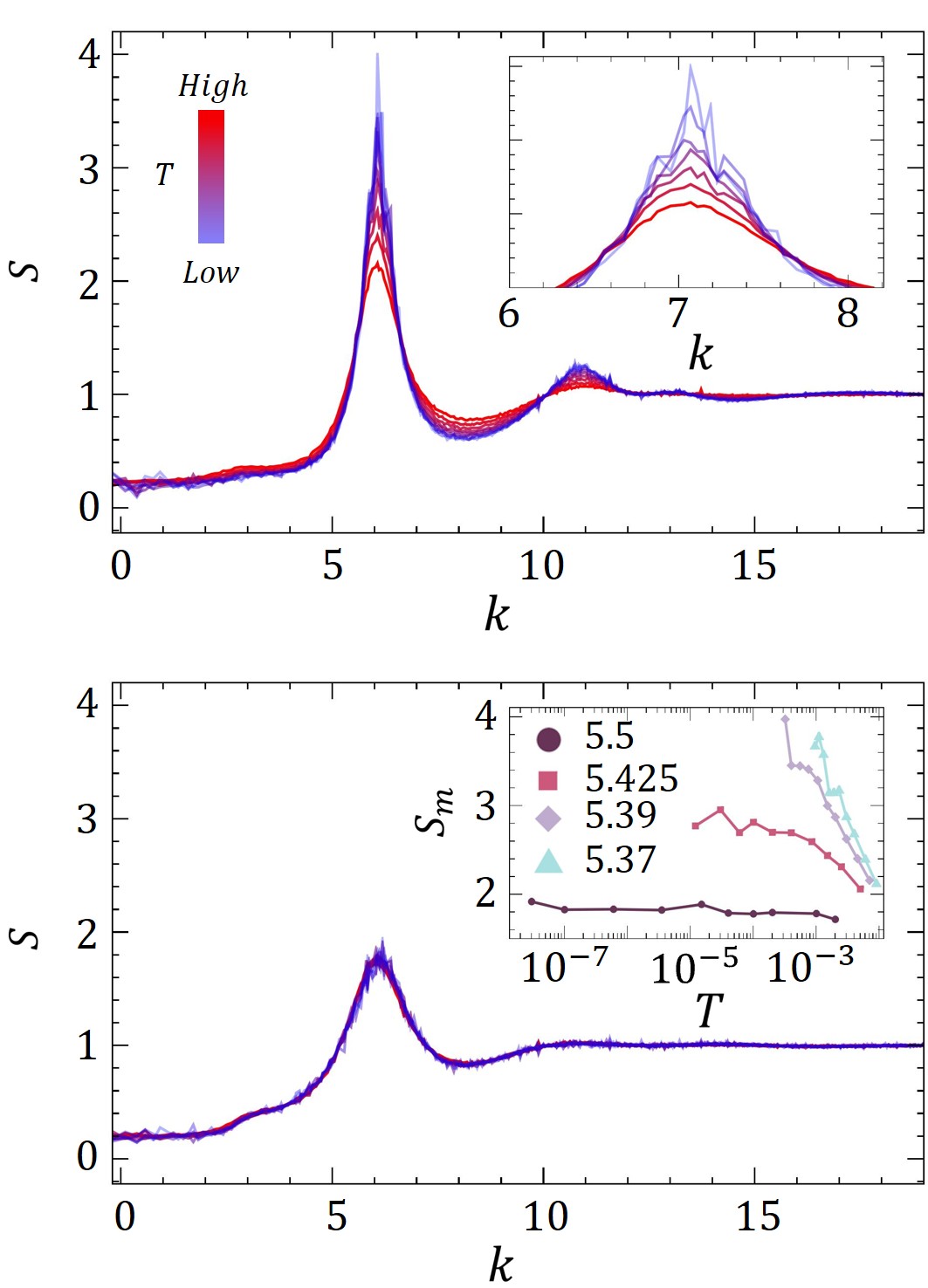}
	\caption{
		\textbf{The temperature dependence of peak height of static structure factor is controlled by $s_0$.}
		Static structure factor $S(k)$ for $s_0=5.39$ (Top) and $s_0=5.50$ (Bottom) at different temperatures (temperatures decrease from red to blue).
		The inset in the top panel magnifies the region corresponding to the first peak in the main panel; the inset in the bottom panel shows the temperature dependence of the first peak height of $S(k)$ for different $s_0$.
	}
	\label{fig:sk}
\end{figure}

\subsection{Structural properties}
\autoref{fig:sk} shows the static structure factor $S(k)$ for two representative values of the control parameter, $s_0=5.39$ (top) and $s_0=5.50$ (bottom), where the main panels display $S(k)$ over the full range $0 \leq k \leq 20$.
We find that the wavevector $k$ corresponding to the first peak of $S(k)$  is essentially independent of both temperature and $s_0$, indicating that the underlying length scale of the packing is fixed.

However, the behavior of the \emph{height} of the first peak, $S_{m}$, strongly depends on $s_0$ (\autoref{fig:sk}, bottom inset).
For $s_0=5.39$, $S_{m}$ grows systematically as temperature decreases  (\autoref{fig:sk}, both insets), resembling the behavior of conventional glassformers.
In contrast, at $s_0=5.5$ the structure factor is nearly temperature independent, with no noticeable change in $S_{m}$ upon cooling.

This latter trend mirrors observations made in simulated vitrimers, a distinct physical system that also exhibits sub-Arrhenius scaling tuned by density \cite{ciarella2019understanding}.
In that system, a crossover from sub-Arrhenius to Arrhenius dynamics was similarly linked to the temperature dependence of the first peak in $S(k)$, a connection further supported by schematic mode-coupling theory.
The parallels between our results and those vitrimer studies suggest that the temperature dependence of the first peak of $S(k)$ may strongly correlate with the character of glassy dynamics across disparate physical models.

\subsection{Mechanical properties}
In order to link the observed $\tau_\alpha$ behavior (\autoref{fig:angelplothmc}) to mechanical properties, we first tried to quantify an intermediate-time plateau shear modulus $G_p$, as we had done for the 2D Voronoi model before \cite{li2025connecting}.
We first calculate the shear stress $\Sigma_{xy}$ at each time point, and compute the time-dependent shear modulus $G(t)$ as the stress autocorrelation function, with statistics accumulated using a multiple-tau correlator method \cite{ramirez2010efficient}.

\begin{figure}[t]
	\centering
	\includegraphics[width=\linewidth]{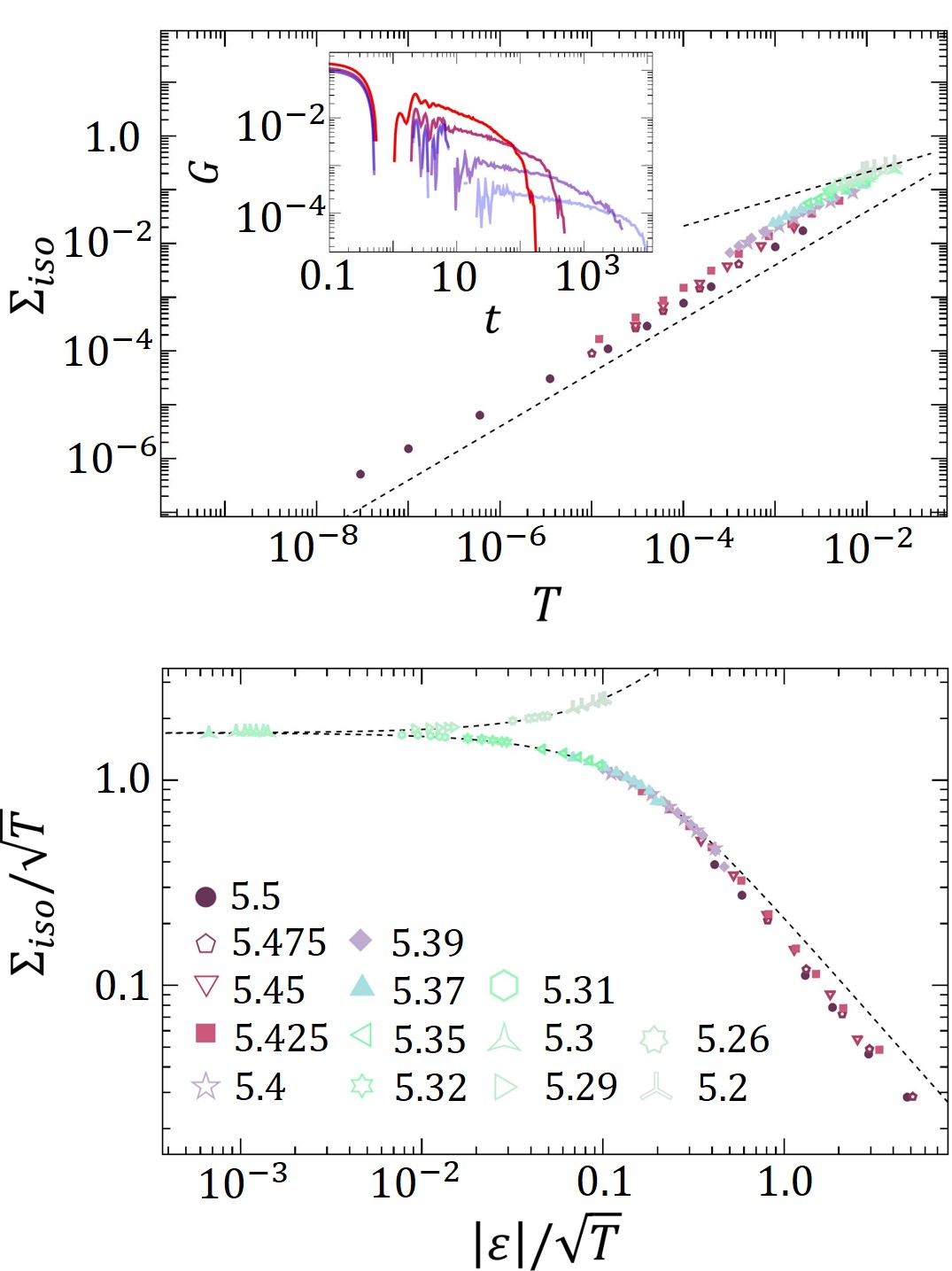}
	\caption{
		\textbf{The isotropic tension increases with temperature and matches the theoretical prediction for generic under-constrained systems.}
		(Top) isotropic tension as a function of temperature for several $s_0$, showing a transition from $T^{1/2}$ to approximately linear $T^1$ scaling.
		Inset: shear relaxation modulus $G(t)$ for $s_0=5.425$, displaying an intermediate-time plateau that decreases monotonically with temperatures (from red to blue, temperature decreases).
		(Bottom) When plotting isotropic tension $\Sigma_{\textrm{iso}}$ versus isotropic strain $\epsilon$, a rescaling by $\sqrt{T}$ leads to a collapse of the curves for the different state points.
		The dashed lines correspond to Eq. \ref{eq:isotropicTension} with parameter values as given in the main text.
	}
	\label{fig:mech}
\end{figure}

In \autoref{fig:mech} (top inset) we plot $G(t)$ for $s_0=5.425$ and several temperatures.
Qualitatively, the behavior matches that of the 2D model, with the expected initial fast relaxation followed by a secondary feature whose magnitude \emph{increases} as the temperature is raised.
This is, however, a notoriously noisy and data-intensive way of extracting mechanical properties \cite{ramirez2010efficient}. For the 3D Voronoi model, the data were too noisy to reliably fit the plateau to a stretched exponential form to extract a precise $G_p$.

To study a proxy for material rigidity, we use isotropic tension $\Sigma_\textrm{iso}$, defined as the trace of the stress tensor.
Indeed, in generic under-constrained systems with fixed connectivity and without shear strain, isotropic tension $\Sigma_{\textrm{iso}}$ is proportional to the infinite-time shear modulus $G$ \cite{lee2024generic,Lee2024a}. It is thus a reliable measure of the system's rigidity while being much more accessible to numerical simulations.
The top panel of \autoref{fig:mech} shows the temperature dependence of $\Sigma_{\textrm{iso}}$ for several values of $s_0$.
As in two dimensions, $\Sigma_{\textrm{iso}}$ displays an unusual temperature dependence: rather than decreasing modestly upon heating, it \emph{increases} monotonically with temperature.
At high temperatures, the tension grows approximately as $T^{1/2}$, while at low temperatures it follows a nearly linear $T^1$ dependence.
This crossover is consistent with what we previously found in the 2D Voronoi model \cite{li2025connecting}, further supporting the robustness of this unusual mechanical response. 

The isotropic tensions we measure are quantitatively consistent with a generic theory for under-constrained materials with fixed connectivity \cite{lee2024generic,Lee2024a}.
Refs.~\cite{lee2024generic,Lee2024a} express the elastic material properties, including $\Sigma_{\textrm{iso}}$, in terms of temperature $T$ and isotropic strain $\varepsilon$ as shown in Eq.~\ref{eq:isotropicTension} in Appendix~B. 
Moreover, isotropic strain $\varepsilon$ can be mapped to the target cell surface area using a length rescaling \cite{Lee2024a}: 
$\varepsilon=1/2\log\left(s_0^\ast/s_0\right)$, where $s_0^\ast$ denotes the athermal transition point.
The bottom panel of \autoref{fig:mech} shows the corresponding scaling plot, which leads to an excellent collapse of curves for different state points $(s_0,T)$.

From a fit of Eq.~\ref{eq:isotropicTension} to our simulation data (dashed line in \autoref{fig:mech} bottom) we obtain the values of athermal transition point $s_0^\ast=5.301\pm0.005$, energetic elasticity $\kappa_E=13.5\pm0.2$, and entropic elasticity $\kappa_S=0.22\pm0.1$ (see Appendix~B for details).
These values of $s_0^\ast$ and $\kappa_S$ lie within the expected ranges: $s_0^\ast$ is slightly below the transition point of 5.41 obtained from infinite-temperature quenches \cite{merkel2018geometrically}, which we attribute to the annealing effect of our finite-temperature simulations. 
For the same reason we also expect the effective transition point $s_0^\ast$ to depend weakly on both $s_0$ and $T$.
Finally, Ref.~\cite{lee2024generic} shows that $\kappa_S=(N_{1st}^\ast-2)/6N$ in dimensionless units, where $N_{1st}^\ast$ is the number of first-order zero modes (see the precise definition in Ref.~\cite{Lee2024a}). Given that the 3D Voronoi model has $3N$ degrees of freedom (the cell positions) but $2N$ constraints (area and volume terms in Eq.~\ref{eq:energyFunctional}), it is reasonable to assume $N_{1st}^\ast=N$ up to terms of order one, and thus $\kappa_S\approx 1/6$, close to what we obtain here.
It is somewhat harder to make clear predictions for $\kappa_E$ based on past work, since we have used a mix of several cell types here.

\begin{figure}
	\centering
	\includegraphics[width=\linewidth]{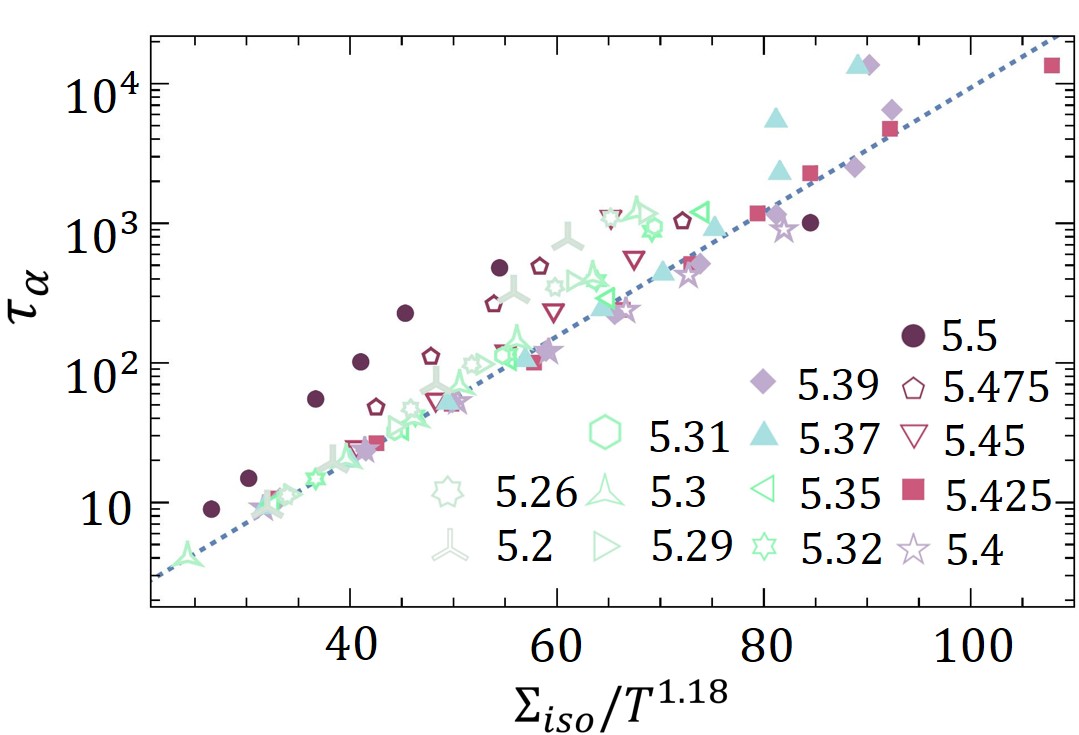}
	\caption{
		\textbf{The mechanics and dynamics are directly connected.}
		$\log(\tau_\alpha)$ is plotted against scaled isotropic tension $\Sigma_{\textrm{iso}}/T^{1.18}$ for many different $s_0$.
	}
	\label{fig:connectionplot}
\end{figure}

Finally, we ask whether these elastic properties alone can explain the observed behavior of $\tau_\alpha$.
Specifically, we ask whether structural relaxation can be understood as the crossing of energy barriers of height $\sim\Delta E$, which is expected to scale with a plateau shear modulus $G_p$ (or, similarly, with isotropic tension $\Sigma_{\textrm{iso}}$) in ``shoving''-like models \cite{li2025connecting,dyre2012instantaneous}.
In \autoref{fig:connectionplot} we plot $\log \tau_\alpha$ against the scaled isotropic tension $\Sigma_{\textrm{iso}}/T^{1.18}$. Just as in the 2D Voronoi model~\cite{li2025connecting}, we find than a temperature exponent different from $-1$ does a much better job of collapsing the data, although we do not have a theoretical explanation for where this originates from.
Across these state points, the majority of data points collapse onto a straight line, and we have confirmed that an ansatz in which $\log \tau_\alpha \sim A \Sigma_{\textrm{iso}}/T$ for a temperature-independent constant $A$ does a much worse job of collapsing the data.
There are some systematic deviations at the lowest temperatures for low-$s_0$ simulations, where $\tau_\alpha$ grows more rapidly than the scaling indicated by the data collapse.
However, the most severe systematic deviations occur for large $s_0$. We hypothesize that this may be a result of the measured $\tau_\alpha$ being dominated by the unusually large relaxation time of beta processes, as discussed earlier (\autoref{fig:fs} top).  

Taken together, like in the 2D Voronoi model \cite{li2025connecting}, also in the 3D version, much of the sub-Arrhenius behavior of $\tau_\alpha$ can be understood from an increase of mechanical rigidity with temperature.

\section{Discussion}

In this work, we have demonstrated that the 3D Voronoi model exhibits tunable sub-Arrhenius relaxation dynamics governed by its unusual mechanical properties.
By moving from the isostatic 2D limit to an extensively under-constrained 3D system, our results establish that anomalous glassy dynamics are not a dimensional artifact or a consequence of 2D long-wavelength fluctuations, but rather a robust, generic property of under-constrained materials.

The most striking evidence for this is the direct connection between the system's structural relaxation and its mechanical rigidity.
Simple elastic models of glass formation posit that the activation energy for structural relaxation $\Delta E$ scales with a local, intermediate-time shear modulus $G_p$ \cite{dyre2012instantaneous,puosi2012communication}.
Because extracting $G_p$ in 3D is notoriously noisy, we utilized the isotropic tension $\Sigma_{\textrm{iso}}$ as a robust proxy for rigidity.
Remarkably, plotting $\log \tau_\alpha$ against the scaled isotropic tension, $\Sigma_{\textrm{iso}} / T^{1.18}$, yields a collapse of data across a broad range of state points.
While a standard shoving model would predict a $T^{-1}$ scaling, the empirical $T^{-1.18}$ dependence confirms that the sub-Arrhenius behavior is primarily driven by a mechanical activation barrier that stiffens at higher temperatures, with the anomalous exponent likely reflecting additional temperature-dependent structural correlations.

Our analytical and structural analyses further illuminate the nature of this under-constrained glassy state.
By applying an analytical theory for the elastic properties of thermal, under-constrained systems to our numerical data, we identified a rigidity transition at $s_0^\ast=5.301$.
This value is lower than the previously reported athermal transition ($s_0^\ast=5.41$) derived from infinite-temperature quenches, likely reflecting the annealing effects of our finite-temperature simulations, though it remains above the crystalline Weaire-Phelan limit ($s_0=5.288$) \cite{Weaire1999}.

Moving away from this transition point dramatically alters the dynamical landscape.
In the high-$s_0$ regime, the energy landscape becomes exceptionally shallow; $\beta$-relaxation processes become progressively slower and compete with the measured $\alpha$-relaxation time, while dynamical heterogeneity is largely suppressed.
Furthermore, the static correlations in this regime captured by the first peak of the structure factor mirror the atypical temperature independence seen in simulated vitrimers.
Conversely, as $s_0$ decreases toward the rigid regime, the model recovers a more conventional glass-forming phenomenology with pronounced dynamical heterogeneities and growing static structural peaks upon cooling, all while maintaining a universal underlying $\alpha$-relaxation process.
These qualitative features are in excellent agreement with the behavior reported in the 2D Voronoi model \cite{li2025connecting}.

Ultimately, the 3D Voronoi model provides a minimal, highly controllable platform for exploring the origins of anomalous glassy dynamics. 
Our findings broaden the class of disordered materials known to exhibit highly unusual glassy phenomenology.
Comparing the mechanics and dynamics of the Voronoi model with other systems (such as vertex models \cite{sussman2018anomalous}, vitrimers \cite{ciarella2019understanding}, or low-density hard spheres \cite{berthier2009compressing}) may soon allow us to uncover a unifying physical framework for sub-Arrhenius disordered dynamics.

\section*{Conflicts of interest}
There are no conflicts to declare.


\section*{Acknowledgements}

This material is based upon work supported by the National Science Foundation under Grant No. DMR-2143815.
This research used the Delta advanced computing and data resource which is supported by the National Science Foundation (award OAC 2005572) and the State of Illinois. Delta is a joint effort of the University of Illinois Urbana-Champaign and its National Center for Supercomputing Applications.

\appendix

\section*{Appendix}

\subsection*{Appendix A: Bayesian inference to extract $\tau_\alpha$}\label{app:Bayesian}

Extracting the relaxation time from stretched exponential fits, $A \exp\!\left[-(t/\tau_\alpha)^{\beta}\right]$, to $F_s(t)$ is notoriously delicate. 
Because directly fitting this functional form requires a nonlinear least-squares optimization that is highly sensitive to the initial parameter guess, we employed a standard Bayesian inference approach to determine the full posterior distributions of the fitting parameters.
Within this framework, we introduce a hierarchical structure that couples parameters across all state points. Crucially, motivated by the excellent collapse of the $\alpha$-relaxation master curves shown in \autoref{fig:fs}, we enforce common behavior in $A$ and $\beta$ while allowing the temperature-dependent relaxation times $\tau_\alpha$ to vary independently.

In order to estimate parameters for the relaxation of the self-intermediate scattering function, we performed Bayesian inference using the \texttt{PyMC} probabilistic programming framework and the No-U-Turn Sampler (NUTS) algorithm \cite{pymc2023}.
The hierarchical model for the stretched exponential decay treated the amplitude $A$ and the stretching exponent $\beta$ as global parameters, while the relaxation time $\tau_\alpha$ was allowed to vary with temperature and $s_0$.
All model parameters were constrained to their physical domains by reparameterization.
The stretching exponent $\beta\in(0,1)$ and the amplitude $A\in(0,1)$ were represented as sigmoid transforms of unconstrained Normal random variables,
\begin{align}
	q_\beta &\sim \mathcal{N}(0,1), & \beta &= \mathrm{sigmoid}(q_\beta), \\
	q_A     &\sim \mathcal{N}(0,1), & A     &= \mathrm{sigmoid}(q_A).
\end{align}
This choice places weakly informative priors on $\beta$ and $A$, centered around $0.5$ with broad support over $(0,1)$.

For the temperature-dependent relaxation times $\tau_\alpha$, we worked with logarithmic variables,
\begin{align}
	\log \tau_\alpha \sim \mathcal{N}(\mu_\tau, \sigma_\tau^2),
\end{align}
with $\mu_\tau=\log(10)$ and $\sigma_\tau=5$, providing wide support over several decades in $\tau$.
The exponential transform ensures $\tau_\alpha>0$.
The likelihood is constructed using a Gaussian probability with the mean and variance computed directly from the distribution of the 100 independent $F_s$ trajectories at each state point.

Posterior sampling was carried out using four independent chains, with 2,000 warm-up iterations and 2,000 sampling iterations per chain, yielding approximately 8,000 posterior draws.
Convergence was assessed using the potential scale reduction statistic $\hat{R}\leq 1.01$ and by inspecting effective sample sizes.
Posterior predictive checks confirmed that the fitted model reproduced both the decay shape and the variance structure of the measured $F_s(t)$.
From the resulting posterior distributions we extracted credible intervals for $\tau_\alpha(T)$.
This hierarchical Bayesian approach produces stable and statistically consistent fits, quantifies uncertainty in a principled way, and exploits shared physical information across the entire dataset.

\subsection*{Appendix B: Prediction of elastic properties of 3D Voronoi tissue}
\label{app: generic under-constained}

In generic under-constrained systems without shear strain, the isotropic tension can be expressed as \cite{lee2024generic,Lee2024a}:
\begin{equation}\label{eq:isotropicTension}
    \Sigma_{\textrm{iso}} = \frac{\kappa_E}{2}\left(\varepsilon+\sqrt{\varepsilon^2+\frac{4\kappa_ST}{\kappa_E}}\right),
\end{equation}
where $\kappa_E$ is the \emph{energetic} isotropic bulk modulus and $\kappa_S$ is the \emph{entropic} isotropic bulk modulus.
Moreover, $\varepsilon$ is the isotropic strain, measured with respect to the athermal transition point; note that athermal under-constrained systems can be driven across the floppy-rigid transition by applying isotropic strain \cite{Merkel2019,Lee2022,lee2024generic,Lee2024a}.
In other words, varying isotropic strain $\varepsilon$ is equivalent to varying the target cell surface $s_0$ \cite{Merkel2019,Lee2022,Lee2024a}. According to the non-dimensionalization in Ref.~\cite{Lee2024a}:
\begin{equation}\label{eq:ep}
	\varepsilon = \frac{1}{2}\log\frac{s_0^\ast}{s_0}
\end{equation}
with $s_0^\ast$ being the athermal transition point.

Strictly speaking, the analytical results in Refs.~\cite{lee2024generic,Lee2024a} were derived for systems lacking a state of self-stress (SSS) in the athermal floppy regime. 
In contrast, cellular models like the 3D Voronoi model strictly possess a SSS, a direct mathematical consequence of the total system volume being fixed (i.e., the sum of all cell volumes is invariant). 
However, based on previous analysis of the 3D Voronoi model \cite{Merkel2019}, the macroscopic constraint restricting the $N$ volume degrees of freedom largely decouples from the surface area constraints. 
Consequently, the ``trivial'' bulk modulus arising from the total volume constraint drops out of the dominant low-energy mechanics, allowing us to safely apply the formalism such that $\kappa_E$ in Eq.~\ref{eq:isotropicTension} corresponds to the isotropic bulk modulus contributed exclusively by the cell surface tension terms.



\balance


\bibliography{citations} 
\bibliographystyle{rsc} 
\end{document}